\documentclass[aps,prl,twocolumn,groupedaddress,superscriptaddress,amsmath,amssymb,
]{revtex4-2}

\usepackage{graphicx, float}
\usepackage{dcolumn}
\usepackage{bm}

\begin{document}

\title{Multiple droplets on a conical fiber: formation, motion, and droplet mergers}

\author{Carmen L. Lee}
\thanks{These two authors contributed equally}
\affiliation{Department of Physics and Astronomy, McMaster University, 1280 Main Street West, Hamilton, Ontario, L8S 4M1, Canada} 
\author{Tak Shing Chan}
\thanks{These two authors contributed equally}
\affiliation{Mechanics Division, Department of Mathematics, University of Oslo, 0316 Oslo, Norway}
\author{Andreas Carlson}
\affiliation{Mechanics Division, Department of Mathematics, University of Oslo, 0316 Oslo, Norway}
\author{Kari Dalnoki-Veress}
\email[]{dalnoki@mcmaster.ca}
\affiliation{Department of Physics and Astronomy, McMaster University, 1280 Main Street West, Hamilton, Ontario, L8S 4M1, Canada} 
\affiliation{UMR CNRS Gulliver 7083, ESPCI Paris, PSL Research University, 75005 Paris, France.}

\begin{abstract}
Small droplets on slender conical fibers spontaneously move along the fiber due to capillary action. The droplet motion depends on the geometry of the cone, the surface wettability, the surface tension, the viscosity, and the droplet size. Here we study with experiments and numerical simulations, the formation, spontaneous motion, and the eventual merger, of multiple droplets on slender conical fibers as they interact with each other. The droplet size and their spacing on the fibre is controlled by the Plateau-Rayleigh instability after dip-coating the conical fiber. Once these droplets are formed on the fiber, they spontaneously start to move. Since droplets of different size move with different speeds, they effectively coarsen the droplet patterning by merging on the fiber. The droplet merging process affects locally the droplet speed and alters the spatiotemporal film deposition on the fiber. 
\end{abstract} 

\maketitle

\section{Introduction}

In Nature, several species of plants~\cite{guo2015experimental, ju2012multi,pan2016upside} and animals~\cite{zheng2010directional, bai2010directional,parker2001water} have evolved to form slender conical structures. One purpose of these conical structures is to transport liquid drops along the cone, either toward or away from the organism. For instance, the spines on a cactus can create a surface preferential to fog condensation, with droplets spontaneously travelling from the tip of the cone to the base~\cite{Wang,Malik,Chen2018, shanahan2011behaviour}. The hydrophobic hairs on the legs of a water strider work to expel errant water droplets off of the insect, allowing the strider to stay on the surface of the lake or pond~\cite{zhang2015self,wang2015self}. Beyond these two examples, there are many more organisms and structures that exhibit this behaviour. Several studies have been conducted to examine the efficacy of droplet transport along these conical structures, and have served to motivate the development of bio-inspired fog harvesting devices~\cite{WangFibers,xu2016fogcollect,cao2014facile,heng2014branched}.

\begin{figure}[t]
\centering
  \includegraphics[width =0.48\textwidth]{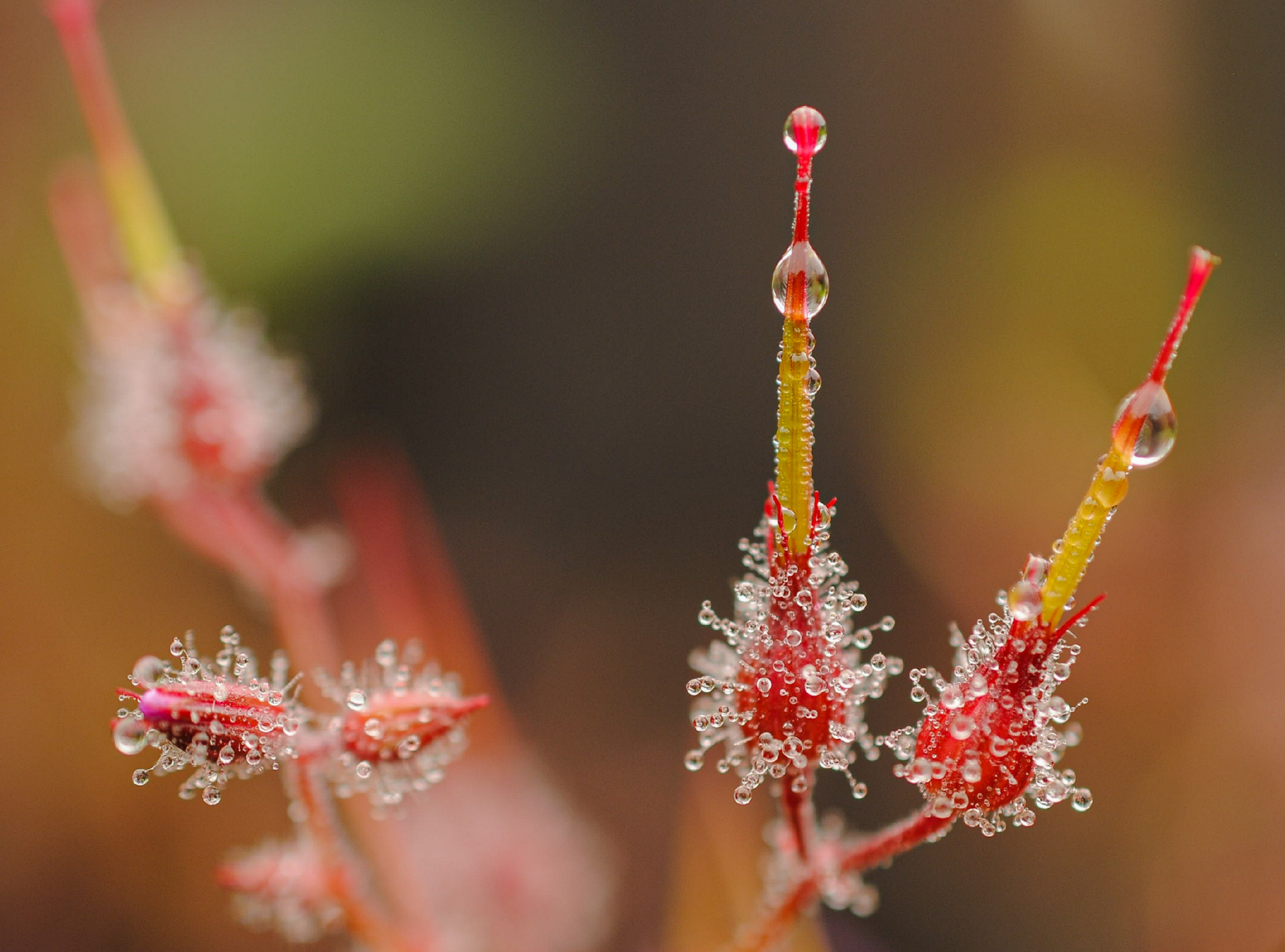}
  \caption{Multiple dew drops covering \textit{Geranium robertianum} fruit. Image credit: Calum Davidson, Aberdeenshire, Scotland}
\label{herb}
\end{figure}

Droplet motion can be driven by several factors. For droplets below the capillary length, motion is driven by the surface tension of the droplet to maximize contact with the fiber when the droplet wets the fiber~\cite{Lorenceau,mccarthy2019dynamics,Fournier,chan2020a,zhang2015self}. As the characteristic size of the droplet becomes larger than the capillary length, the droplet may then be either propelled~\cite{GiletSliding,JiBertozzi,Li,weyerfibernetwork} or stalled~\cite{chou2011equilibrium} by gravity, and motion depends on the orientation of the conical structure and motion can be induced in these systems by temperature gradients or coatings~\cite{hou2013temperature, yarin2002motion}. The motion of a single droplet on a slender conical structure has been studied and characterized extensively, with an excellent understanding of the driving forces and viscous dissipation in the droplet~\cite{Chan2020, ChanLee2021, Fournier, Li, Lorenceau, lv2014substrate, vanhulle2021barreltoclam,gupta2021}. Furthermore, previous works have examined the formation of multiple droplets via the Plateau-Rayleigh instability (PRI) of a film of liquid on a cylindrical fiber~\cite{haefner2015cap,roe1975wetting}. In addition, studies have examined wetting and coalescence of droplets on a fiber network with particular interest in droplet capture in textiles~\cite{safavi2021capture, wang2021hysteresis, zhang2020inkdrop, labbe2019aerosol, WeyerArray}  However, there has been little work on the formation and interaction between multiple droplets as they traverse along the fiber beyond qualitative observations~\cite{Lorenceau, haefner2015cap, Lu}, despite being prevalent in nature as illustrated in Figure~\ref{herb}; a photograph of the \textit{Geranium robertianum} fruit
 covered with dew drops.
 
Here we examine the formation of multiple viscous liquid droplets along a conical fiber created by  the PRI. We then follow the motion of the droplets as they move along the fiber, propelled by capillary forces and the interaction between droplets when they meet and coalesce. We rationalize the experimental observations with numerical simulations of the droplet flow based on a thin film model. The experiments and simulations are found to be in excellent agreement.

\section{Experimental Methods}
We fabricate ideal conical fibers with a smoothly changing radial gradient from borosilicate glass capillary tubes (1 mm outer diameter, World Precision Instruments)~\cite{Fournier, ChanLee2021}. The capillaries are locally heated and pulled with a constant force using a micropipette puller (Narishige PN-30). The resulting shape is a smoothly tapering conical glass fiber with the base of the fiber being the diameter of the original capillary tube (1 mm), and the tip being tens of microns in diameter.

\begin{figure}[h]
\centering
  \includegraphics[width =0.4\textwidth]{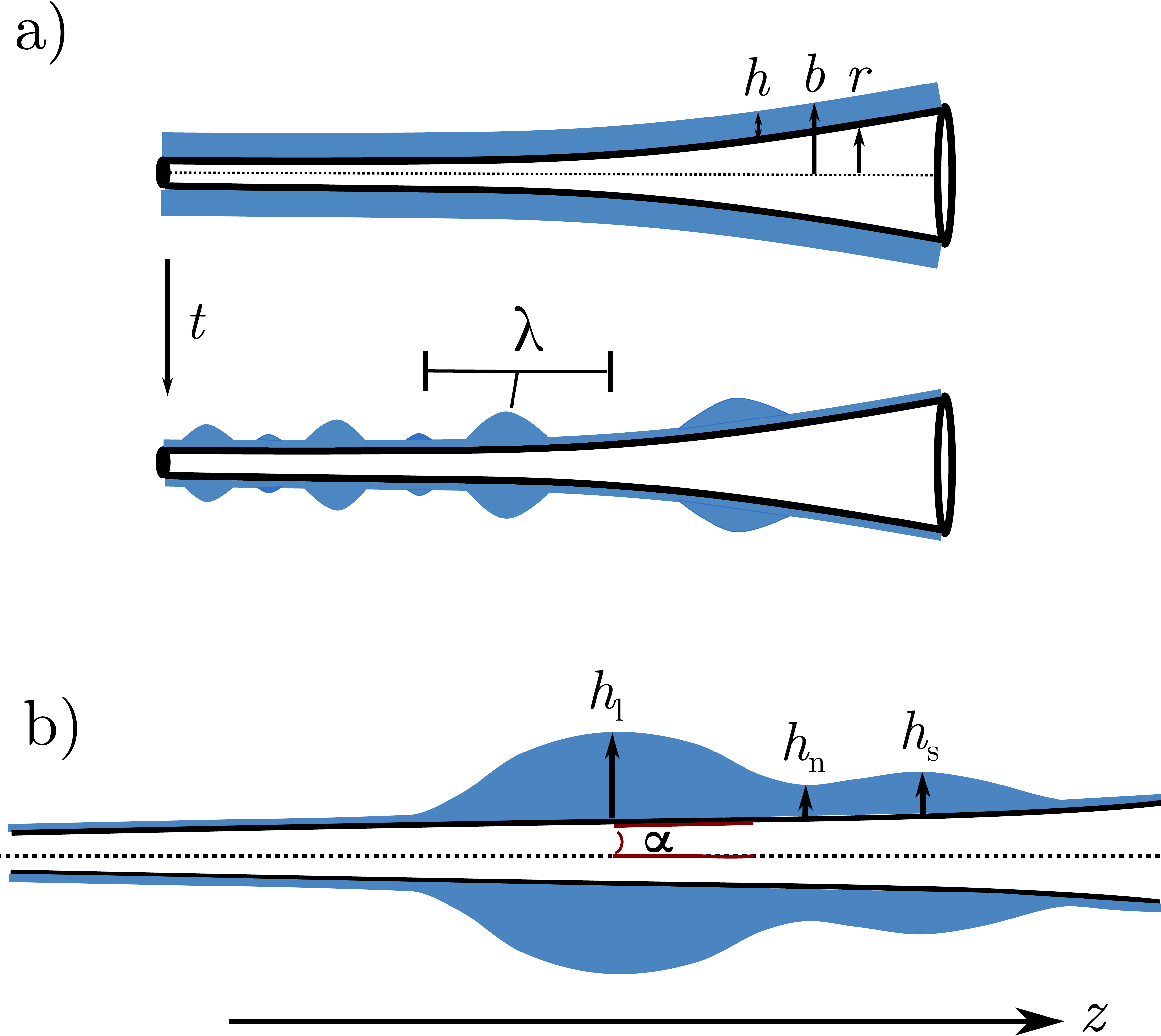}
  \caption{a) A schematic diagram of a coated glass fiber with radius $r(z)$ and film thickness $h$. After some time, $t$, the film breaks up into droplets via the Plateau-Rayleigh instability, where the thickness from the axi-symmetric centre to the liquid/air interface, $b$ varies along the fiber. b) A schematic of two droplets merging on a conical fiber characterised by angle $\alpha$, with height $h_\textrm{l}$ for the large droplet, $h_\textrm{s}$ for the small droplet, and $h_\textrm{n}$ for the minimum of the liquid-air interface between the droplets, as a function of horizontal position $z$.}
\label{schematic}
\end{figure}

A silicone oil (vinyl terminated polydimethylsiloxane) with kinematic viscosity $\eta =$ 5000 cSt was used to create droplets in these experiments. The silicone oil is used because it is non-volatile, non-hygroscopic, and is totally-wetting on glass, with a surface tension $\gamma = 22$~mN/m. Droplets are placed on the conical fiber in two ways. The first uses the Landau-Levich-Derjaguin (LLD) film deposition~\cite{Koulago1995}, where the fiber is inserted into a reservoir of fluid and then pulled away at a constant speed, leaving behind a uniform fluid film on the fiber. The thickness of the film depends on the speed that the reservoir moves, with a film thickness that increases with increasing speed. The resulting film is unstable via the PRI, and will break up into droplets. Figure~\ref{schematic}a) shows a schematic diagram of this process, with relevant geometric parameters labelled. The patterning of droplet formation along the fiber is controlled by the thickness of the LLD film.

The second method begins with creating droplets in the method described in figure~\ref{schematic}a) on a different fiber. The resulting droplets are transferred by bringing one of the droplets into contact with the fiber~\cite{Fournier}. This method allows for the precise placement of individual droplets along the fiber. A droplet can be placed at the tip of the fiber and allowed to completely move to the base of the fiber to deposit a thin film to act as a lubrication layer for subsequent droplets. To examine droplet coalescence, we take advantage of the dependence of droplet speed on the size of the droplet. A large droplet moves more quickly toward the base of the fiber compared to a smaller droplet~\cite{Fournier}. We place a smaller droplet closer to the base of the fiber, and a larger droplet is placed closer to the tip allowing for the larger droplet to catch up to the smaller droplet, coalesce, and merge.

With the fiber held horizontal, the droplet motion is recorded from above using an optical microscope. Additionally, a reference image of the bare glass pipette is taken before the experiment begins. From these images, the fiber radius,  cone angle, and droplet heights, can be extracted as a function of position along the fiber, and as a function of time $t$. Image analysis was done with an in-house Python script.

\section{Numerical simulations}
To model the motion of the droplets, we describe the flow inside the liquid by the Stokes equations and impose a no-slip boundary condition at the solid substrate, and no-shear stress at the liquid-air interface. Assuming that both the slope of the liquid-air interface and the cone angle $\alpha$ are small, we apply the lubrication approximation~\cite{Batch67,Oron97} to the Stokes equations and obtain the thin film equation for a conical geometry.  A detailed derivation of the thin film equation can be found in~\cite{chan2020a}. The axisymmetric liquid-air interface profile is given by $h=h(\zeta,t)$, defined as the distance between the interface and the substrate, as a function of the distance from the vertex of the cone $\zeta$ and time $t$.  We note that $\zeta$ measures along the surface of the cone and is related to the experimentally measured $z$ as $z =\zeta \textrm{cos}(\alpha)$ and for small angles $\zeta \approx z$. The evolution of the liquid-air interface is described by the thin film equation and driven by the capillary pressure gradients $\partial p/ \partial \zeta$ which reads~\cite{chan2020a,chan2020b},
\begin{eqnarray}\label{lac1}
\frac{\partial h}{\partial t}+\frac{1}{\zeta\alpha+h}\frac{\partial}{\partial \zeta} \left(M \frac{\partial p}{\partial \zeta}\right)=0,
\end{eqnarray}
where the mobility $M=M(h,\zeta,\alpha)$ is expressed as

\begin{equation}
\begin{aligned}
M(h,\zeta,\alpha)=&\frac{\zeta^4\alpha^4}{2\mu}\Bigg\{\frac{1}{8}\Bigg[3\left(1+\frac{h}{\zeta\alpha}\right)^4  \\
      & -4\left(1+\frac{h}{\zeta\alpha}\right)^2+1\Bigg]\\
      &-\frac{1}{2}\left(1+\frac{h}{\zeta\alpha}\right)^4\ln\left(1+\frac{h}{\zeta\alpha}\right)\Bigg\}.
      \label{G}
\end{aligned}
\end{equation}

The capillary pressure gradient in the liquid generates the flow and the pressure $p=p(\zeta,t)$ reads

\begin{equation}\label{cur}
p=-\gamma\Bigg\{\frac{\frac{\partial^{2} h}{\partial \zeta^{2}}}{\left[1+\left(\frac{\partial h}{\partial \zeta}\right)^{2}\right]^{3/2}}-\frac{1-\alpha \frac{\partial h}{\partial \zeta}}{(r\alpha 
	+h )\left[1+\left(\frac{\partial h}{\partial \zeta}\right)^{2}\right]^{1/2}}\Bigg\},
\end{equation}
where the expression is simplified for $\alpha\ll 1$~\cite{chan2020a,chan2020b}. Eq.~(\ref{lac1}) and (\ref{cur})  are discretized by linear elements and numerically solved with a Newton solver by using the open source finite element code FEniCS\cite{logg2012automated}, additional details about the numerical approach are found in~\cite{chan2020b}. The initial condition is a droplet smoothly connected to a pre-wet film of thickness $\epsilon$.  At the two boundaries ($\delta\Omega$) of the numerical domain we impose $h(\delta\Omega,t)=\epsilon$ and $p(\delta\Omega,t)=\gamma/R(\delta\Omega)$, where $R(\delta\Omega)$ is the radius of the cone at the boundaries. We note that only the droplet volume $V$ is important and the initial droplet shape does not affect the results.

\section{Results and Discussion}
\subsection{Plateau-Rayleigh instability on a conical fiber}
\begin{figure}[]
\centering
  \includegraphics[width =0.48\textwidth]{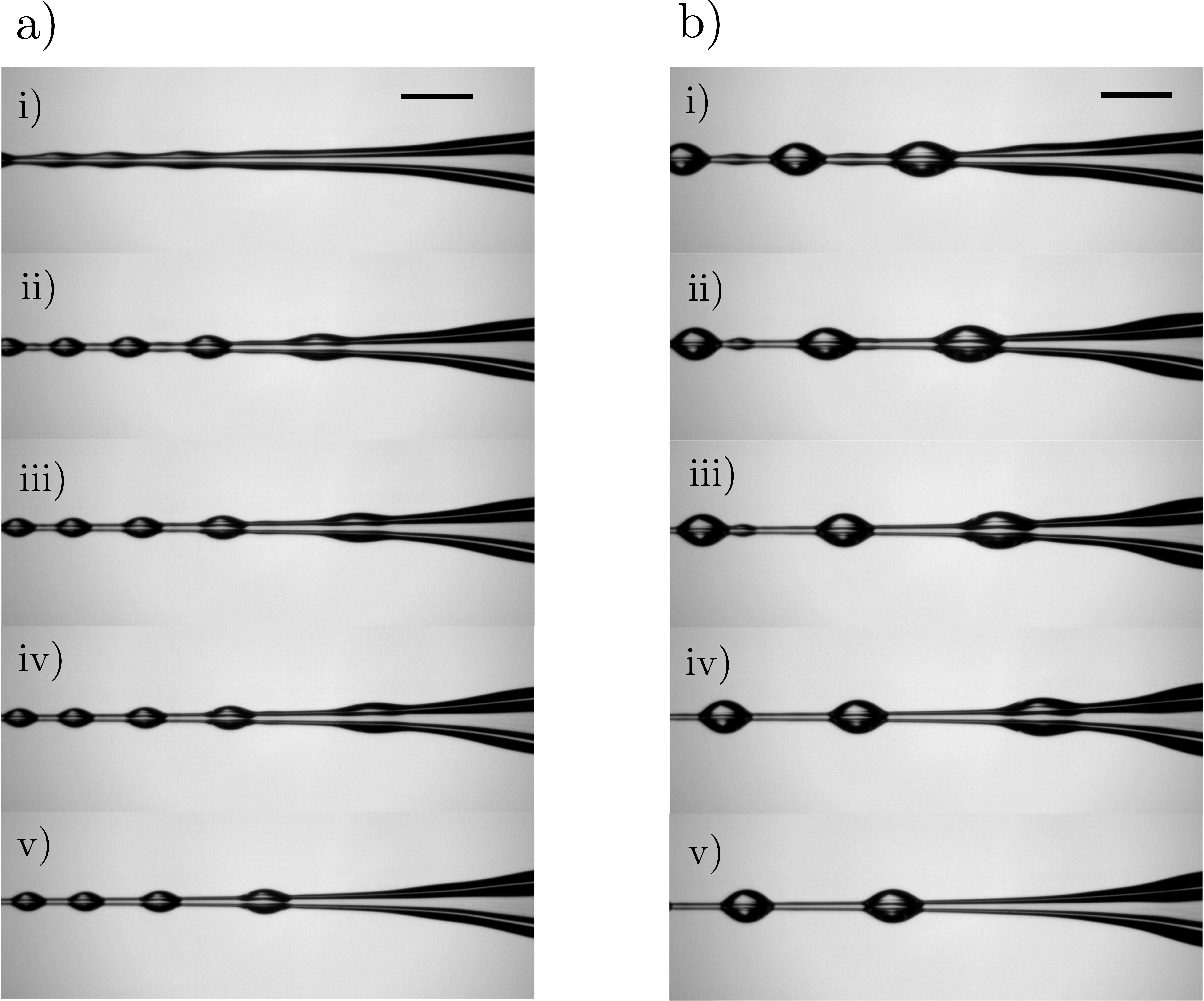}
  \caption{Optical microscope images of a fiber coated with silicone oil with film thicknesses a) $h$ = 53 $\pm$1 $\mu$m b) $h$ = 88 $\pm$1 $\mu$m. Panels i) -v) show snapshots of the break up of the coating film and the motion of the droplet from left to right along the fiber. The film thickness in a) shows the formation of monotonically increasing droplets, where the film thickness in b) breaks up into principle and satellite droplets. Scale bar indicates 1000 $\mu$m.}
\label{pri}
\end{figure}

A cylinder of liquid will break up into droplets to minimize the surface area according to the PRI. The size and spacing of the droplets increases with the initial radius of the cylinder: in the case of a liquid film with thickness $h$, coating a solid cylinder with radius $r$, the sum of the radius of the cylinder and liquid film, $b = h+r$, dictates the wavelength. However, a conical fiber does not have a constant radius along the length of the fiber, and hence the wavelength between droplets is not constant. As such, the patterning of droplets along a conical fiber will depend on the thickness of the film, and the location along the fiber. To test the effect of the film thickness, the reservoir was pulled at different constant velocities to deposit films of different thicknesses. Figure~\ref{pri} shows examples of the break up of the coating film for two different values of $h$: the thinner film, $h=53~\mu$m, shown in Fig.~\ref{pri}a) was pulled out of the reservoir more slowly than the thicker film with $h=88~\mu$m shown in Fig.~\ref{pri}b). For the thinner film [Fig.~\ref{pri}a)], droplets form in panel i) with monotonically increasing sizes along the fiber, and they move progressively toward the base of the fiber in the subsequent frames ii)-v). In figure~\ref{pri}b), a different patterning appears, where larger droplets form [panel i)] with small satellite droplets in between. At later times, [panels ii) and iii)] the larger droplets catch up with the satellite droplets and coalesce, resulting in the large droplets moving toward the end of the fiber [panels iv) and v)].

\begin{figure}[h]
\centering
  \includegraphics[width =1\columnwidth]{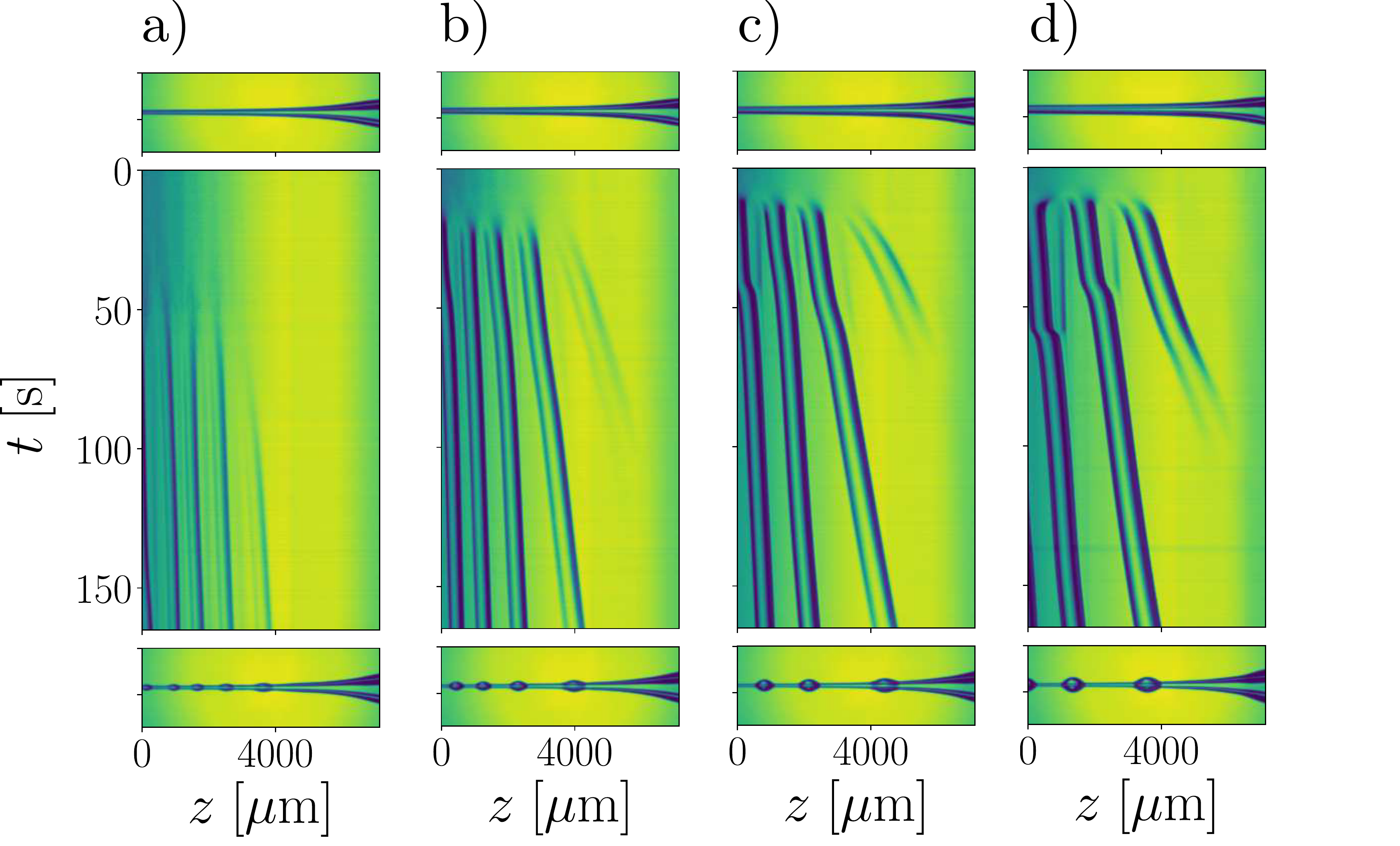}
  \caption{Experimental images of a fiber coated with silicone oil. The top row corresponds to the fiber coated with film thickness, a) $h= 29 \pm1\ \mu$m,  b) $53 \pm1 \ \mu$m , c) $77 \pm1 \ \mu$m, and d) $88 \pm 1 \ \mu$m at the initial state $t=0$. The middle panel shows the central slice of each image, stacked to reveal the the break up of the film into droplets and their resulting motion, giving the position $z$ of the droplets as a function of time $t$. The final row shows the patterning of the droplets at the end of the observation time.}
\label{greenplot}
\end{figure}

We can visualize the motion of the droplets by taking a row of pixels along the axis of the fiber for each time-step of the optical microscopy images. This row is stacked for each image in the time-series to show the migration of droplets as a function of time in figure~\ref{greenplot}. In figure~\ref{greenplot} the same fiber is coated with four different film thicknesses ranging from $h = 29\pm1\ \mu$m to  $h = 88 \pm 1 \ \mu$m with the top row corresponding to the initial film at $t=0$. The dark bands in the center panel correspond to the beginning and end of each droplet. Using these images we track the droplet locations along the fiber. In figure~\ref{greenplot}c) and d) one can observe the principle droplets moving with a faster velocity (steeper slope) compared to the smaller satellite droplets. Once the small and large droplets come into contact, we see the trace of the smaller droplet disappear when it merges with the large droplet. The final droplet pattern is shown in the bottom row.

\begin{figure}[ht]
\centering
  \includegraphics[width =0.48\textwidth]{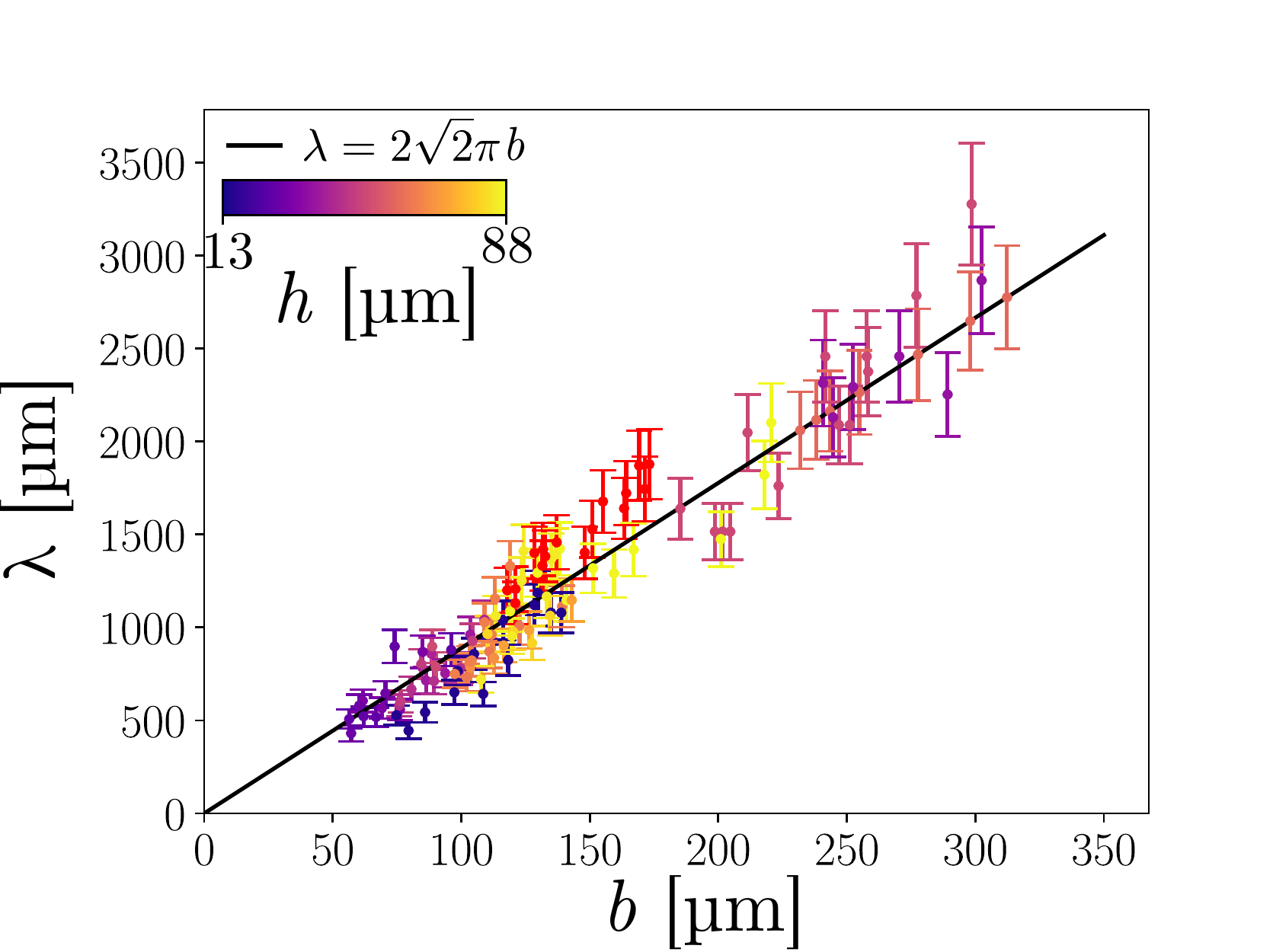}
  \caption{Spacing between droplets ($\lambda$) produced from the Plateau-Rayleigh instability on a conical fiber. Different colours refer to different liquid film thicknesses, while the horizontal axis, the distance from the axi-symmetric centre to the liquid/air interface, $b$, depends on the position along the conical fiber, $z$. }
\label{wavelength}
\end{figure}

The relationship between wavelength of droplet formation and the radius of the fiber and coating is shown in figure~\ref{wavelength}. For various pipettes and film thicknesses, the initial wavelength between droplets is shown with different experiments indicated by different colours. The conical shape of the fiber results in different time constants along the length of the fiber, meaning that the droplets appear at different times and immediately move after forming, therefore the wavelength was defined when the droplet first appears with a pronounced minima on either side. To find the wavelength, we locate the position of the nearest local maxima (\textit{i.e.} the neighbouring principle droplets) on either side of a droplet, which corresponds to twice the wavelength. The data are shown in figure~\ref{wavelength}. The solid line shows the relationship $\lambda = 2\pi\sqrt{2}b$, where $\lambda$ is the wavelength, $b$ is the distance from the axis of symmetry about the length of the fiber and the liquid/air interface, \textit{i.e.} a combination of the fiber radius $r$ with the thickness of the coating film $h$~\cite{InstabJets}. The wavelengths found using this method show excellent agreement with the theory and provides insight into using conical fibers to passively control the patterning of droplets.

For sufficiently large film thicknesses, the PRI occurs such that satellite droplets form between the principal droplets, which can be seen in figure~\ref{pri}b) panel i). Previous experiments and models on the motion of a single droplet on a conical fiber show droplet speed along a fiber increases with droplet volume~\cite{Lorenceau, Fournier}, due to the increased volume reducing the resistance to flow. In addition, the larger the fiber radius the slower the droplets move due to the increased contact with the fiber. Fournier \textit{et al.}~\cite{Fournier} showed the importance of the lubricating layer on the speed of the droplet: the absence of a lubricating film drastically slowed down the motion of the droplets in comparison to those droplets with a lubricating film, indicating that equivalent droplets on the same fiber move faster with a thicker precursor film compared to a thinner film.  Based on these previous experiments, as the droplets move along the fiber, the larger main droplets catch up to the smaller satellite droplets that are closer to the base of the fiber. The two droplets merge to form a large droplet and change the initial distribution of drops created by the PRI.

\subsection{Coalescing droplets}
\begin{figure}[]
\centering
  \includegraphics[width =0.4\textwidth]{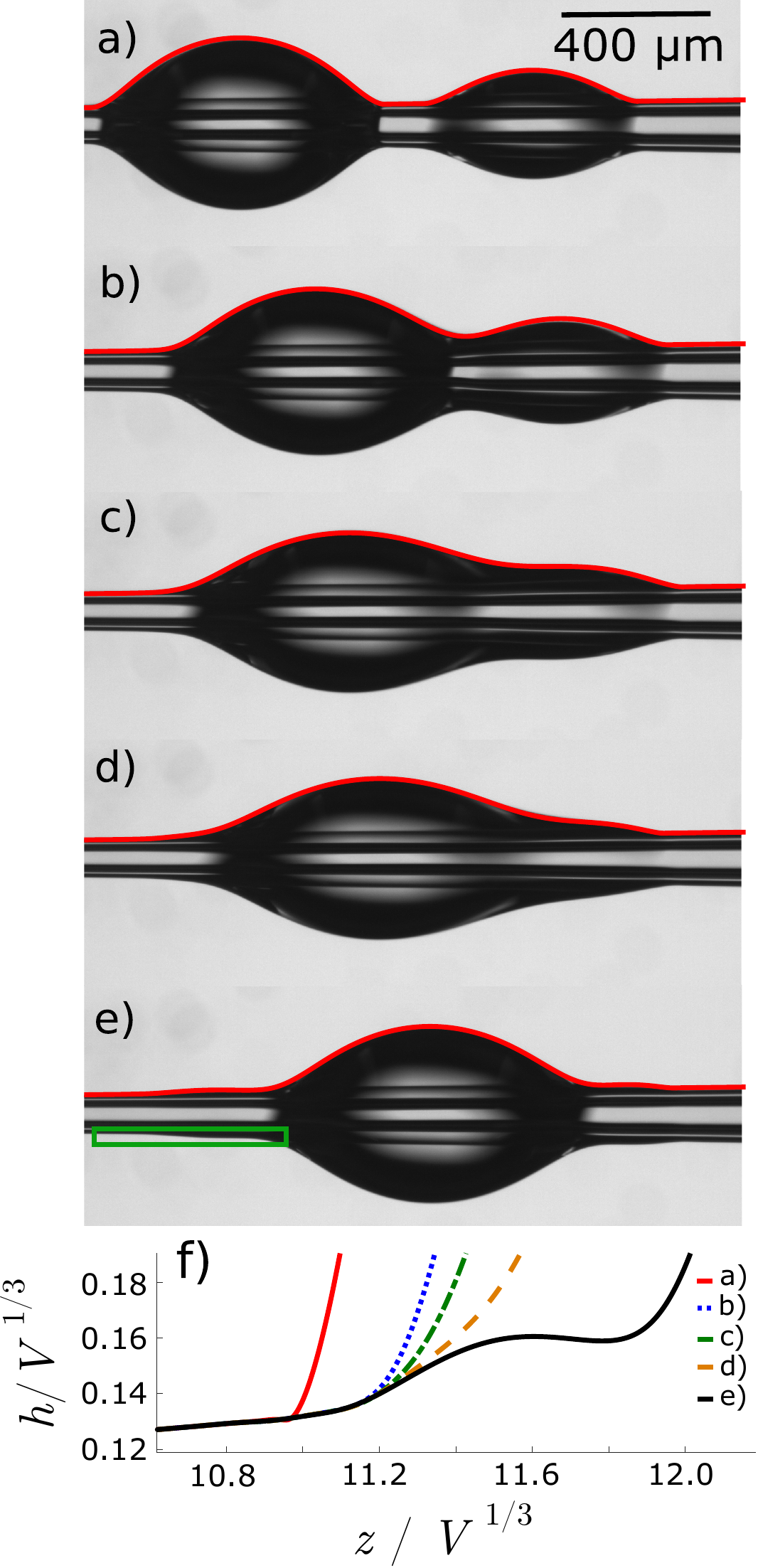}
  \caption{Images showing the main steps in the coalescence process of two droplets ($V$ =  0.0191 $\pm$0.0002 mm$^3$ and 0.0050 $\pm$0.0003 mm$^3$) on a conical fiber ($\alpha = 0.673 \pm0.002^\circ$). a) separate droplets. b) two droplets when first in contact with one another. c) Flow of the smaller, higher pressure droplet backwards into the larger droplet. d) Acceleration of the large droplet toward the film left behind by the small droplet. e) A final droplet, the volume of the two initial droplets. f) The dimensionless profiles of the deposited film $h/V^{1/3}$ as a function of $z/V^{1/3}$ at different times (from a)-e)) in the receding region of the big droplet (indicated by the green box in e)). A thicker film is deposited behind the droplet after the two droplets merge.}
\label{merging}
\end{figure}

To examine the coalescence process in more detail, individual droplets were carefully placed on the fiber using the second droplet deposition method described in the experimental section. These droplets were imaged at higher magnification compared to the PRI droplets. Images of the merging process are shown in figure~\ref{merging}. In figure~\ref{merging}, we observe five distinct stages: a) The two droplets begin as individual droplets. b) As the larger droplet catches up with the smaller droplet, a thick fluid bridge forms between the two, connecting the two fluid bodies. The smaller Laplace pressure in the bridge causes the liquid in the droplets to flow toward the bridge. c) The bridge thickens as liquid flows into the bridge and the local maximum height of the smaller droplet decreases. d) Remnants of the small droplet leave behind a thick film preceding the larger droplet. e) The droplet reforms into the semi-symmetric shape, and continues moving to the end of the fiber. A thicker film is deposited behind the merged droplet due to the acceleration during merging along with the higher velocity associated with the larger volume of droplet. See figure~\ref{merging}f).

In addition to optical microscopy, figure~\ref{merging} also shows profiles of the merging process matched to the conditions of the experiment as generated by the numerical simulations. Outlined in red, the profiles shown here match exactly the profiles obtained from experiment. An interesting feature that appears in both simulation and experiment is the presence of a thicker film deposited behind the droplet after merging (shown in figure~\ref{merging}e, outlined in green). Focusing on the location outlined in green in figure~\ref{merging}e), we can extract the profile of LLD film left behind by the moving droplet, to show that the increase in speed of the droplet due to merging, deposits a thicker film than a slower moving droplet~\cite{ChanLee2021}.

\begin{figure}[]
\centering
  \includegraphics[width =0.48\textwidth]{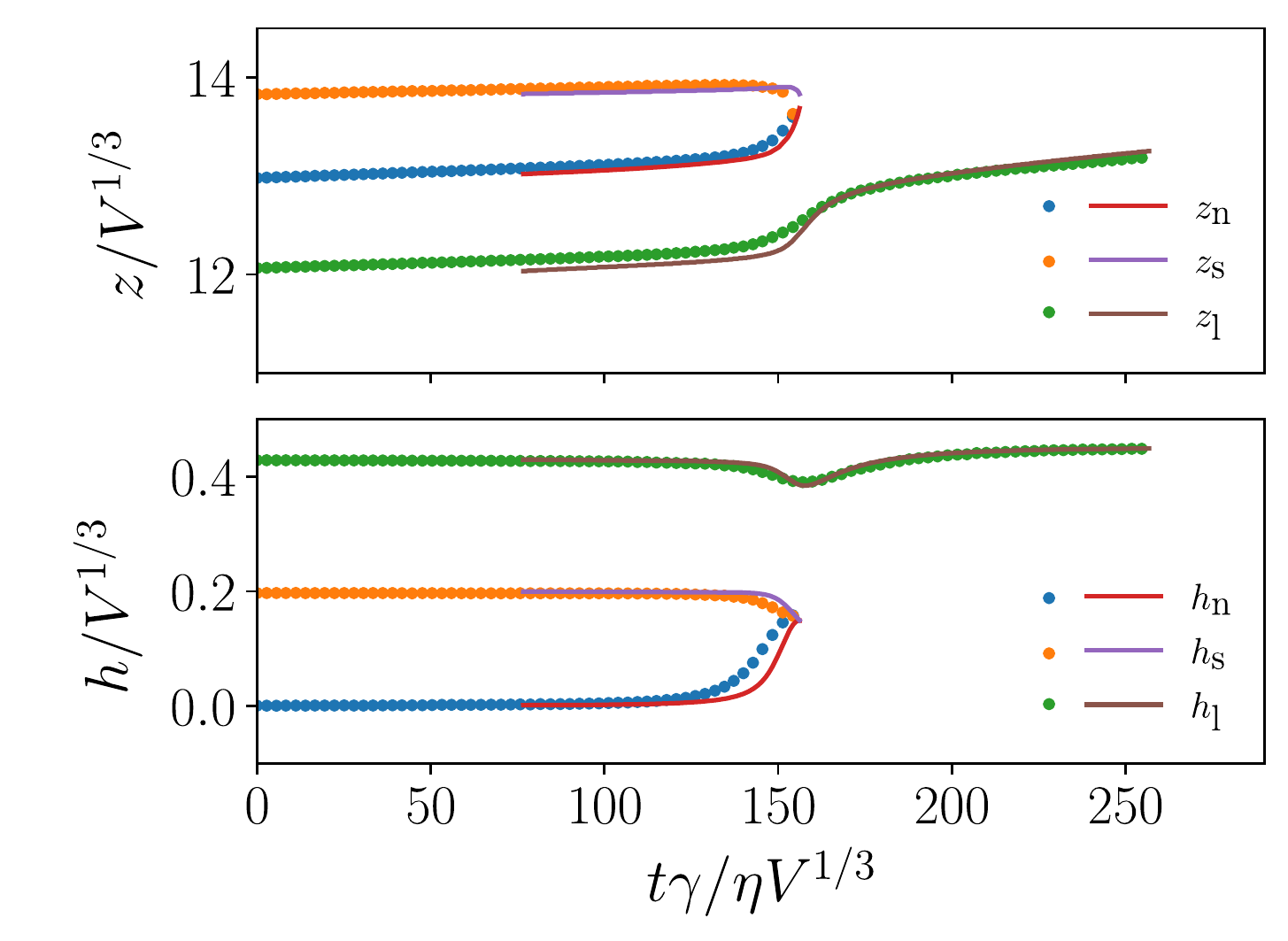}
  \caption{a) The position along the fiber $z$ rescaled with the volume of the large droplet $V$ for the maximum point of the large and small droplets along with the fluid neck as a function of dimensionless time. b) The renormalized height of the liquid/air interface at the local maxima of the large and small droplets $h_\textrm{l}$ and $h_\textrm{s}$, respectively, and the local minima between the two droplets marking the location of the liquid neck $h_\textrm{n}$ are plotted as a function of time $t$. The heights are normalized by the length scale determined by the volume of the large droplet $V$. Experimental values for both subplots are denoted by dots and the matched numerical simulations are denoted by lines.}
\label{expt_theory}
\end{figure}

To further understand the dynamics during droplet merging, we track the position of the droplets. We do this by locating three features on the droplets, the positions and height of two maxima associated with the large ($z_\textrm{l}(t)$, $h_\textrm{l}(t)$) and small droplet ($z_\textrm{s}(t)$, $h_\textrm{s}(t)$), and the position of an intermediate radius ($z_\textrm{n}(t)$, $h_\textrm{n}(t)$), which denotes the location of the local minima in the fluid neck (see figure~\ref{schematic}b). The position along the fiber $z$ is measured relative to the tip of the cone with a constant cone angle.
Figure~\ref{expt_theory}a) shows the position of these three points, normalized by the length scale determined by the volume of the large droplet $V$, as a function of dimensionless time $ t \gamma /\eta V^{1/3}$, and figure~\ref{expt_theory}b) shows the rescaled heights $h/V^{1/3}$ of each of these points as a function of dimensionless time $ t \gamma /\eta V^{1/3}$. In addition to the experimental data, we have matched the droplets with simulations. A slight difference in the experimental data compared to the simulations can be attributed to the fiber having a gently varying cone angle along the length of the fiber, unlike the simulated fiber which has a constant cone angle. Regardless, the agreement between experiment and theory is excellent.

\section{Conclusions}
In this work, we have studied the formation of multiple droplets on a conical fiber through the breakup of a coating film via the Plateau-Rayleigh instability. The patterning of droplet production depends on the gradient of radius of the cone, where the averaged wavelength or droplet spacing depends on the radius of the fiber at the location of the drop. The initial spacing of the droplets can be explained through the relationship predicted by the classic Plateau-Rayleigh instability theory. The droplets are driven by the surface tension to spontaneously move along the fiber, where the motion of the drops depends on the size of the droplet, properties of the liquid and geometry of the fiber. Depending on the local thickness of the cylinder where the droplet is created, smaller satellite droplets may form between larger drops, and with time, the larger droplets catch up with the smaller droplets and merge, changing the droplet pattern. We have captured the main steps of droplet merging, including bridge formation, pressure equalization and reformation into a large droplet. We have modelled the merging of the droplets using the approach of lubrication approximation on a conical geometry for Stokes flow which closely matches the dynamics captured in the experiments. Understanding the motion and the interaction of droplets merging on a conical fiber can give insight into efficient methods of fog harvesting and pattern formation through the Plateau Rayleigh instability also gives insight into the process of self-patterning droplet on fibers.

\begin{acknowledgments}
CLL and KDV gratefully acknowledge the financial support by the Natural Science and Engineering Research Council of Canada. TSC and AC gratefully acknowledge financial support from the Research Council of Norway NANO2021 program (Project No. 301138) and the UiO:LifeScience initiative at the University of Oslo.
\end{acknowledgments}

\end{document}